\documentclass{desyproc}

\begin{document}
\title{Searching for galactic axions through magnetized media: QUAX status report }

\author{{\slshape G. Ruoso$^1$, D. Alesini$^2$, C. Braggio$^{3,4}$, G. Carugno$^{3,4}$, N. Crescini$^{1,4}$, D. Di Gioacchino$^2$, P. Falferi$^{5,6}$, S. Gallo$^{3,4}$, U. Gambardella$^8$, C. Gatti$^2$, G. Iannone$^8$,  G. Lamanna$^9$, C. Ligi$^2$, A. Lombardi$^1$, R. Mezzena$^{6,7}$, A. Ortolan$^1$,
R. Pengo$^1$, C. C. Speake$^{10}$}\\[1ex]
$^1$INFN, Laboratori Nazionali di Legnaro,  Legnaro (PD), Italy\\
$^2$INFN, Laboratori Nazionali di Frascati, Frascati (Roma), Italy\\
$^3$INFN, Sezione di Padova, Padova, Italy\\
$^4$Dip. di Fisica e Astronomia,  Padova, Italy\\
$^5$Istituto di Fotonica e Nanotecnologie, CNR, Povo (TN), Italy\\
$^6$INFN, TIFPA,  Povo (TN), Italy\\
$^7$Dip. di Fisica, Povo (TN), Italy\\
$^8$Dip. di Fisica E.R. Caianiello, Fisciano (SA), Italy and INFN, Sez. di Napoli, Napoli, Italy\\
$^9$Dip. di Fisica and INFN, Sez. di Pisa, Pisa, Italy\\
$^{10}$School of Physics and Astronomy, Univ. of Birmingham, Birmingham, United Kingdom}

\contribID{familyname\_firstname}

\confID{13889}  
\desyproc{DESY-PROC-2017-XX}
\acronym{Patras 2017} 
\doi  

\maketitle

\begin{abstract}
The current status of the QUAX R\&D program is presented. QUAX is a feasibility study for a detection of axion as dark matter based on the coupling to the electrons. The relevant signal is a magnetization change of a magnetic material placed inside a resonant microwave cavity and polarized with a static magnetic field.

\end{abstract}

\section{Introduction}

The QUAX (QUaerere AXion) program explores the feasibility of an apparatus to detect axions as a dark matter component  by exploiting its interaction with the spin of electrons (See \cite{QUAX} and references therein). Due to the motion of the Solar System through the galactic halo, the Earth is effectively moving through the cold dark matter cloud surrounding the Galaxy and an observer on Earth could detect such axion wind. In particular, its effect  on a magnetized material can be described as an effective oscillating rf  field with frequency determined by $m_a$ and amplitude related to $f_a$. Thus, a possible detector for the axion wind can be a magnetized sample with Larmor resonance frequency tuned to the axion mass by means of an external polarizing static magnetic field: e.g. 1.7 T for 48 GHz, corresponding to a 200 $\mu$eV axion mass, in the case of the interaction with the electron spin that is considered hereafter. The interaction with the axion effective field drives the total magnetization of the sample, and so produce oscillations in the magnetization that, in principle, can be detected.
To optimize the detection scheme, the sample is placed inside a microwave cavity. The cavity and the magnetized sample have to be cooled down at ultra-cryogenic temperature to reduce the noise due to  thermal photons.

By using the Lagrangian that describes the interaction  between axion and electron, it is possible to derive the amplitude of the effective magnetic field $B_a \equiv \frac{g_p}{2e}  \boldsymbol{\nabla}a $. Here, $g_p$ is the dimensionless pseudo-scalar coupling, $a$ is the axion field, and $e$ is the electron charge.
Axions represent the best example of non-thermal dark matter candidate. In the following we will suppose that axions are the dominant component.  The axion velocities are distributed according to a Maxwellian, with a velocity dispersion $\sigma_v\approx 270$ km/sec. We can treat the axion as a classical field, and the the effective magnetic rf field associated with the mean axion field has the amplitude and frequency given by  

\begin{equation}
B_a  = 2.0 \cdot 10^{-22} \left(\frac{m_a}{200\, \mu {\rm eV}}\right)    \,\,\,\, {\rm T},\,\,\,\,\, \frac{\omega_a}{2 \pi}  =  48  \left(\frac{m_a}{200\,\mu{\rm eV}}\right)    \,\,\,\, {\rm GHz}, \, 
\label{axionfield}
\end{equation}
respectively,
with a relative linewidth $\Delta \omega_a/\omega_a \simeq 5.2 \times 10^{-7}$.  
Among various axion models, this detection scheme is sensitive to  axions described by the DFSZ model \cite{DFSZ}. 

\section{The experiment}

To detect the extremely small rf field $B_a$ we exploit the Electron Spin Resonance (ESR) in a magnetic sample. In particular, we want to collect the power deposited in the sample by the axion wind due to its interaction with the electron spin. To enhance the interaction we will tune the ferromagnetic resonance of the sample, i.e. the Larmor frequency of the electron, to the mass value of the searched for axion. The sample is placed inside a microwave resonant cavity which is used to reduce the effect of radiation damping and to optimize power collection.
The axion wind will effectively drive the magnetization of the magnetic sample, and this will result in the emission of rf radiation that can  be collected by using an antenna critically coupled to the cavity mode. 
 It is useful to write the expected output power by referring to relevant experimental design parameters. For a magnetic sample of volume $V_s$ and spin density $n_S$ we have
\begin{equation}
P_{\rm out}=\frac{P_{\rm in}}{2}=
3.8 \times 10^{-26} \left(\frac{m_a}{200\, \mu{\rm eV}}\right)^3 \left( \frac{V_s}{100\,\, {\rm cm}^3}\right)
 \left( \frac{n_S}{2 \cdot10^{28} /{\rm m}^3}\right)
 \left( \frac{\tau_{\rm m}}{2 \,\mu{\rm s}}\right)\,{\rm W},
 \label{power}
\end{equation}
where the chosen axion mass $m_a$ is determined by a magnetizing field  $B_0=1.7$ T. $\tau_{\rm m}$ is the characteristic time of the detector system, connected to the microwave cavity and magnetic resonance linewidths.

At the moment we are performing an R\&D to study the feasibility of this proposal. Among the relevant issues that we are working at we can mention: 1) {\bf magnetic material}: it should have a spin density of about 2$\times 10^{28}$/m$^3$ with a ferromagnetic linewidth of about 150 kHz, over a total volume of 100 cm$^3$; 2) {\bf microwave cavity}: a Q factor of the order of 10$^6$, operated in a static magnetic field and housing the magnetic material; 3) {\bf magnetizing field}: provide a source up to a 2 T field with high uniformity and stability, both at the ppm level; 4) {\bf microwave receiver}: linear amplifier are limited to quantum noise, so we foresee the use of a microwave single photon counter; 5) {\bf complete apparatus}: the system has to work around 100 mK, with the noise budget limited only by the thermal photons. Some detail for a few of this topics are given below.

\subsection{The magnetic material}

We are currently working with Yttrium Iron Garnet (YIG), which has a spin density of about $2\times10^{28}$/m$^3$ at room temperature, and slightly higher at cryogenic temperature. Highly polished spheres have a linewidth at room temperature of about 2 MHz. 

\begin{wrapfigure}{r}{0.55\textwidth}
  \begin{center}
    \includegraphics[width=0.55\textwidth]{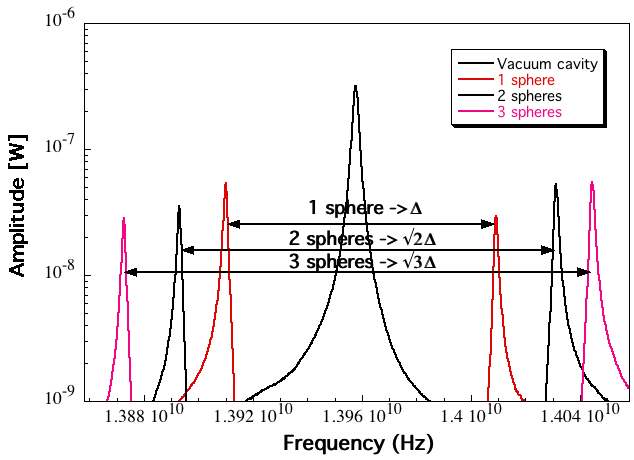}
    \caption{Hybridization measurements with 1, 2 and 3 magnetized YIG spheres coupled to the microwave cavity mode.}
    \label{fig:three}
  \end{center}
\end{wrapfigure}
We studied the temperature dependance of the linewidth, and we confirmed what was suggested in the literature: standard YIG samples are contaminated by rare earth (at the ppm level) and wider linewidths is observed at low temperature. 
High purity YIG spheres instead shows no temperature dependance of the linewidth, down to $\sim$ 10 K. We are now investigating other materials like paramagnets (BDPA, K$_3$CrO$_4$) or ferrimagnets (GaYIG, Lithium ferrite).

Single spheres of YIG are of limited size (a few mm diameter): in order to reach reasonable volumes one has to use many spheres that must be properly aligned to the external polarizing field. The proper alignment can be checked by measuring the coupling between the ferromagnetic resonance of the material and the microwave cavity resonance. In  case of strong coupling, hybridization occurs and the single mode splits in two, with a mode separation proportional to the square root of the total number of active spins.  Figure 1 shows that the mode separation scales   with the square root of the number of spheres as expected. Moreover, no effect on the sample linewidth is evident.

\subsection{The resonant cavity}

\begin{wrapfigure}{r}{0.35\textwidth}
  \begin{center}
    \includegraphics[width=0.45\textwidth, angle=-90]{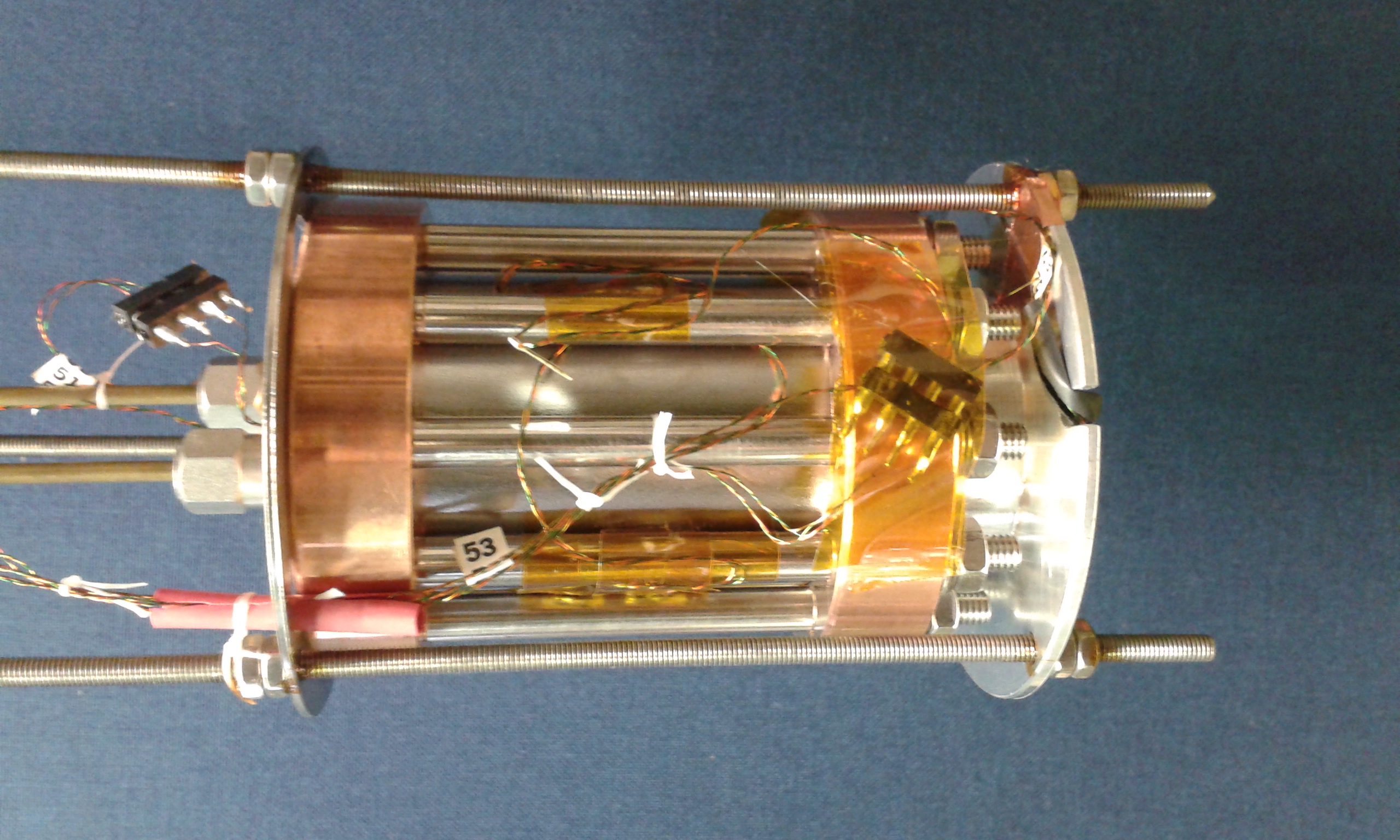}
    \caption{Nb cavity instrumented before the insertion in the LHe criostat.}
    \label{fig:cavity}
  \end{center}
\end{wrapfigure}

The resonant cavity of the QUAX experiment must provide a high quality factor (from few hundred thousands to 1 million) for the TM110 mode at a resonant frequency of about 48~GHz when immersed in a magnetic field of 1.7~T. Moreover, the cavity must allow the penetration of the external magnetic field without deforming it and its resonant frequency must be tunable in a range of about a hundred MHz. For the R\&D phase, the resonant frequency has been reduced to 14~GHz corresponding to $m_a=60\,\mu$eV and to a lower external field of intensity 0.5~T. We aim at reaching these goals through an optimal mechanical design and the proper choice of inner surface materials.
The cavity design foresees conical-shaped end-caps to reduce the current dissipation at interfaces, an asymmetric cross-section of the inner cylinder to remove the mode degeneracy and a frequency tuning obtained through the lateral insertion of longitudinal tuning bars able to shift the frequency of 100~MHz for 2~mm insertion with small deterioration of the quality factor. 
With a 14~GHz copper-cavity, 2 cm inner diameter and 5~cm length, cooled down to 4~K we reached a quality factor about 50,000. To improve quality  we are investigating hybrid cavities, with copper end-caps and lateral surfaces made of bulk or sputtered type II superconductor: Nb (Figure~\ref{fig:cavity}), NbTi film and MgB$_2$. 

A longitudinal cut on the lateral surface allows the magnetic field to penetrate the inner region of the superconducting cylinder. We performed a complete characterization of the Nb cavity with magnetic and RF measurements at 4~K in an external magnetic field up to the critical field, about 0.5~T. The program will continue with tests of NbTi and MgB$_2$~\cite{ref:giunchi} RF cavities at the nominal critical fields of 0.5~T and 1.7~T.

\subsection{The complete apparatus}

We have set-up a reduced scale  apparatus to start the evaluation of  noise performances of the QUAX detector. A copper cylindrical  microwave cavity (resonant frequency 14 GHz) is equipped with a YIG sphere and immersed into a superconducting solenoid. By using a home made ultra stable current generator (at about 30 amps) a 0.5 T field sets the ferromagnetic resonance of the YIG at the cavity resonance frequency. The hybrid modes are then sensed with a critically coupled wire antenna;  the antenna output is amplified by an HFET low noise cryogenic pre-amplifier, followed by a 
room temperature amplifier. The signal is then down converted with a mixer and sampled with a 2 MHz fast ADC. We effectively measure a 1 MHz window centered at one of the two  resonant hybrid modes of the system cavity plus YIG. The cavity, the magnet and the pre-amplifier are housed inside a liquid helium cryostat and the working temperature is about 5 K. We have seen that the rms noise coincides with the expected value due to the thermal background, indicating that no extra noise is added by the material. After integration for about an hour, the analysis of the power fluctuations gave us  the limit sensitivity of about $10^{-22}$ W. For the current configuration this correspond to a limit in the axion effective magnetic field of about $10^{-16}$ T.

\section*{Acknowledgments}

We would like to thank E. Berto, F. Calaon, M. Iannarelli,  G. Pileggi and M. Tessaro for technical help.


\begin{footnotesize}

\end{footnotesize}



\begin{thebibliography}{99}
%
\bibitem{QUAX}
R. Barbieri {\it et al.}, Phys. Dark Univ. {\bf 15}, 135 (2017).
\bibitem{DFSZ}
M. Dine, W. Fischler, and M. Srednicki,   Phys. Lett.   B   {\bf 104},  199 (1981);
M. Dine, W. Fischler, and M. Srednicki,  Nucl. Phys. B  {\bf 189},  575 (1989);
A.R. Zhitnitsky, Soviet J. Nucl. Phys.  {\bf 31},  260 (1980). 

\bibitem{ref:giunchi} G. Giunchi {\it et al.},
Supercond. Sci. Technol. {\bf 20}, L16-L19 (2007).


\end{thebibliography}
\end{document}